\documentclass[%
reprint,
superscriptaddress,
showpacs,preprintnumbers,
amsmath,amssymb,
aps,
pra,
longbibliography
]{revtex4-2}
\usepackage{color}
\usepackage{graphicx}
\usepackage{bm}
\usepackage{mathrsfs}
\usepackage{dsfont}
\usepackage{ulem}
\usepackage{physics}
\usepackage[colorlinks,allcolors=blue]{hyperref}
\usepackage{mathrsfs}

\makeatletter

\newcommand{\Rmnum}[1]{\expandafter\@slowromancap\romannumeral #1@}
\makeatother

\begin{document}

\title{Experimental Demonstration of Efficient High-dimensional Quantum Gates with Orbital Angular Momentum}

\author{Yunlong Wang}
\affiliation{Shaanxi Key Laboratory of Quantum Information and Quantum Optoelectronic Devices, School of Physics of Xi'an Jiaotong University, Xi'an 710049, China}

\author{Shihao Ru}
\affiliation{Shaanxi Key Laboratory of Quantum Information and Quantum Optoelectronic Devices, School of Physics of Xi'an Jiaotong University, Xi'an 710049, China}

\author{Feiran Wang}
\email{feiran0325@mail.xjtu.edu.cn}
\affiliation{School of Science of Xi’an Polytechnic University, Xi'an 710048, China}

\author{Pei Zhang}
\email{zhangpei@mail.ustc.edu.cn}
\affiliation{Shaanxi Key Laboratory of Quantum Information and Quantum Optoelectronic Devices, School of Physics of Xi'an Jiaotong University, Xi'an 710049, China}

\author{Fuli Li}
\email{flli@xjtu.edu.cn}
\affiliation{Shaanxi Key Laboratory of Quantum Information and Quantum Optoelectronic Devices, School of Physics of Xi'an Jiaotong University, Xi'an 710049, China}

\date{\today}

\begin{abstract}
Quantum gates are essential for the realization of quantum computer and have been implemented in various types of two-level systems.
However, high-dimensional quantum gates are rarely investigated both theoretically and experimentally even that high-dimensional quantum systems exhibit remarkable advantages over two-level systems for some quantum information and quantum computing tasks.
Here we experimentally demonstrate the four-dimensional X gate and its unique higher orders with the average conversion efficiency 93\%. All these gates are based on orbital-angular-momentum degree of freedom of single photons.
Besides, a set of controlled quantum gates is implemented by use of polarization degree of freedom.
Our work is an important step towards the goal of achieving arbitrary high-dimensional quantum circuit and paves a way for the implementation of high-dimensional quantum communication and computation.
\end{abstract}

\maketitle

\section{Introduction}
In quantum communication and computation, quantum states are used for coding, carrying, storing and processing information.
Because of its quantum features such as superposition and unitary transformation of quantum states, many novel things emerge possible, such as speed-up quantum computing for some NP problems \cite{walther2005experimental,o2007optical,ghosh2020quantum}, absolutely secure communications \cite{ursin2007entanglement,liao2017long,zhu2017experimental,yin2017satellite,bhaskar2020experimental}, and simulations for quantum many-body systems \cite{kim2010quantum,gerritsma2010quantum,cirac2012goals,georgescu2014quantum,childs2018toward,kokail2019self,hu2019quantum,endo2020variational}.

The qubit, a two-level quantum system and the quantum counterpart of the classical qubit, constitutes an elementary unit of various quantum information and computation protocols.
Apart from qubit, qudit (i.e. high-dimensional quantum systems) can also be used for quantum information tasks. It has been shown that qubit can provide higher information density coding \cite{walborn2006quantum,dixon2012quantum,willner2015optical}, improve security in quantum communication \cite{zhang2008secure,wang2020high}, simplify the implementation of quantum logic circuit \cite{lanyon2009simplifying}, and inspire novel quantum imaging techniques \cite{lloyd2008enhanced,simon2012two}.
In addition,  high-dimensional quantum systems demonstrate superior channel capacity\cite{bechmann2000quantum}, stronger violation of locality\cite{collins2002bell,dada2011experimental,cao2015multiuser}, and higher noise tolerance\cite{ecker2019overcoming}.

The realization of arbitrary unitary transformation operations to quantum states is crucial for quantum information processing. An arbitrary unitary transformation operation on a single qubit can be realized by the Hadamard gate and the phase-shift gate \cite{nielsen2002quantum}.
Unitary transformations of a single qubit and the controlled NOT gate of two qubits are universal for quantum computers \cite{barenco1995elementary}.
Similar to the case of qubits, the generalized Pauli gates $X$ (cyclic transformation) and $Z$ (state-dependent phase) are proved to be universal for qudit operation, from which all $d$-dimensional Weyl operators can be constructed \cite{babazadeh2017high}.

\begin{figure*}[!t]
  \centering
  \includegraphics[width=0.85\linewidth]{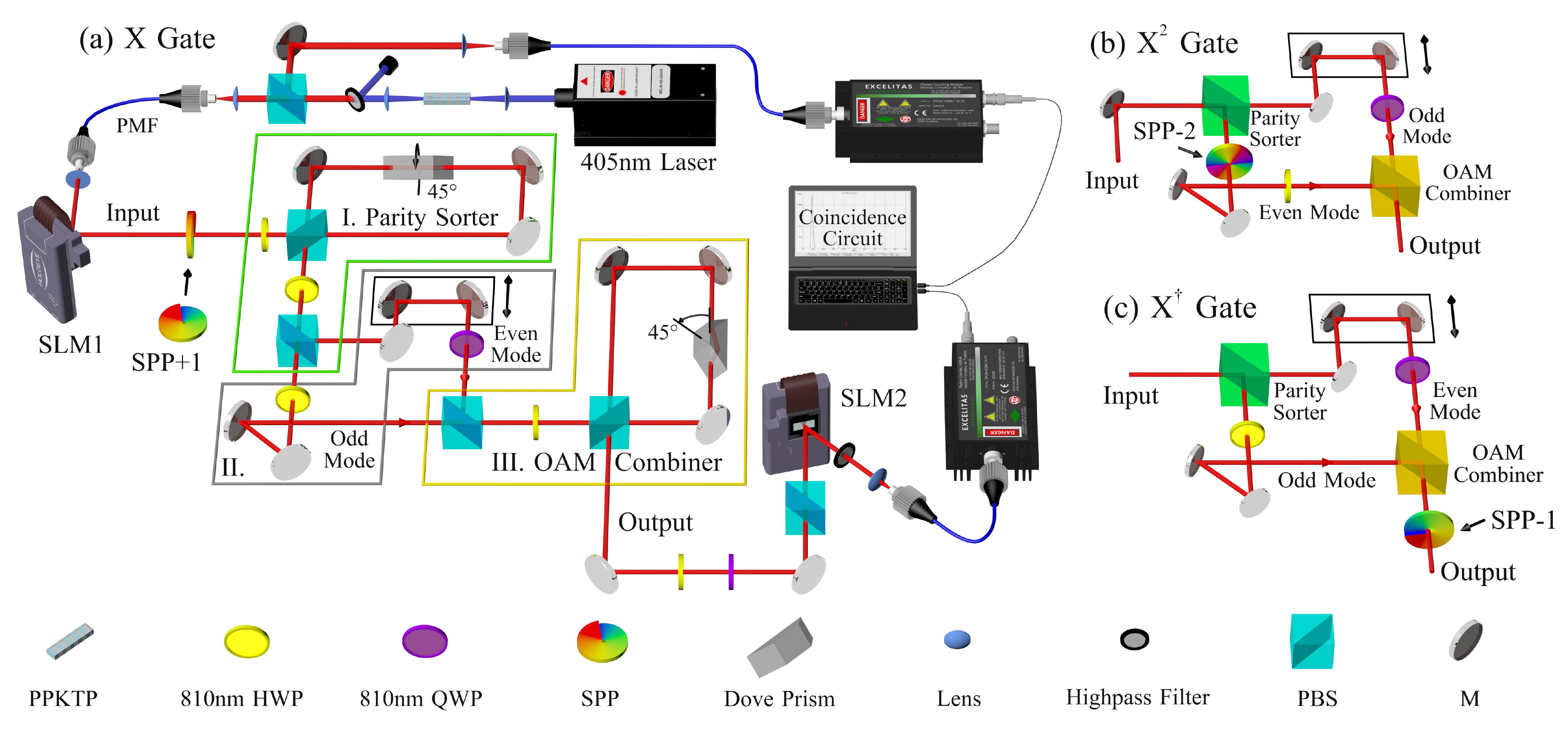}
  \caption{Experimental setup for the four-dimensional cyclic gates. A 405 nm diode laser pumps a type-II ppKTP crystal, and generates orthogonal polarization correlated photon pairs. The idler vertical polarization photon is used for heralding the signal horizontal polarization photon. After passing through a single-mode polarization-maintaining fiber (PMF) and a spatial light modulator (SLM1), the signal photon is modulated into an OAM superposition state. The signal photon goes through the half-wave plate, and its horizontally polarized  state is changed to the diagonally polarized one. Then the photon is reflected or transmitted at the polarization beam splitter (PBS). The horizontally polarized photon undergoes the X-transform operation meanwhile the vertically polarized photon remains unchanged. Eventually, the two paths of the photon are recombined at PBS. The photon is projectively detected through SLM2 and single mode fiber.}
  \label{setup}
\end{figure*}

The temporal and spatial structure of photons provides a natural multi-modal state space in which quantum information can be encoded and processed. Transverse modes of light are discrete and inherently span a high-dimensional Hilbert space.
Specifically, Laguerre-Gaussian (LG) modes of light carry the orbital angular momentum (OAM) $\ell \hbar$  $(\ell=0,1,2,...)$ and constitute a set of orthogonal and complete bases. The LG modes are often used in current experiments for implementing quantum information and computation in a high-dimensional Hilbert space \cite{erhard2018twisted,cozzolino2019high,shen2019optical,wang2020vectorial,erhard2020advances}.
Recently some of high-dimensional quantum gates based on OAM \cite{babazadeh2017high,brandt2020high} and time-frequency of photons have been reported \cite{imany2019high}.  Several high-dimensional quantum gates encoded in the full-field mode structure of photons are experimentally implemented with the phase modulation of spatial light modulator (SLM) in Ref.~\cite{brandt2020high}. Due to the costs incurred by every reflection upon the SLM, the final loss is around $25\%$ which limits the single-photon conversion efficiency in their work. Therefore, efficient methods for control, manipulation and unitary transformation of LG modes still remain a challenging task.

In this work, multiple Sagnac-type and Mach-Zehnder-type polarization-dependent interferometers are cascaded to maintain the state coherence and stability of the whole system for the cyclic OAM operations.
Since an extra parity sorter is inserted as an OAM mode combiner, compared with the previous studies \cite{babazadeh2017high}, we eliminate the additional half of the photon loss meanwhile ensuring the stability of the setup.
In this manner, we achieve the conversion efficiency of the four-dimensional $X$ gate and its integer powers above $ 93\% $ compared to the scheme before.
Additionally, by optimizing SLM patterns to produce the desired input state in our experiment, the overall efficiency is raised.
Combined with the degree of freedom (DoF) of polarization, we have realized polarization-controlled four-dimensional quantum logic gates.
The present work is a step forward to the realization of high-dimensional quantum computation and information processing.

\section{The Implementation of High-dimensional Cyclic Gates}
A qudit is considered as a quantum version of $d$-ary digits whose state can be described by a vector in the $d$-dimensional Hilbert space $\mathscr{H}_d $.  This Hilbert space is spanned by a set of mutually orthonormal basis vectors $\left\{\left| 0\right\rangle,\left| 1\right\rangle,\left| 2\right\rangle,…,\left| d-1\right\rangle \right\}$. In high-dimensional quantum computation, qubit  can be replace by qudit as the basic computational element. The state of a qudit is transformed by qudit gates. The Pauli matrices $\sigma_{x,y,z}$ are basis blocks which construct various unitary transformations of single qubits. The generalized $d$-dimensional Pauli matrix $X$ for qudit is defined as
\begin{eqnarray}
X = \left( {\begin{array}{*{20}{c}}
0&0& \cdots &{\begin{array}{*{20}{c}}
0&1
\end{array}}\\
1&0& \cdots &{\begin{array}{*{20}{c}}
0&0
\end{array}}\\
0&1& \cdots &{\begin{array}{*{20}{c}}
0&0
\end{array}}\\
\begin{array}{l}
 \vdots \\
0
\end{array}&\begin{array}{l}
 \vdots \\
0
\end{array}&\begin{array}{l}
 \ddots \\
 \cdots 
\end{array}&{\begin{array}{*{20}{c}}
 \vdots & \vdots \\
1&0
\end{array}}
\end{array}} \right).
\end{eqnarray}
\begin{figure}[b]
  \centering
  \includegraphics[width=0.7\linewidth]{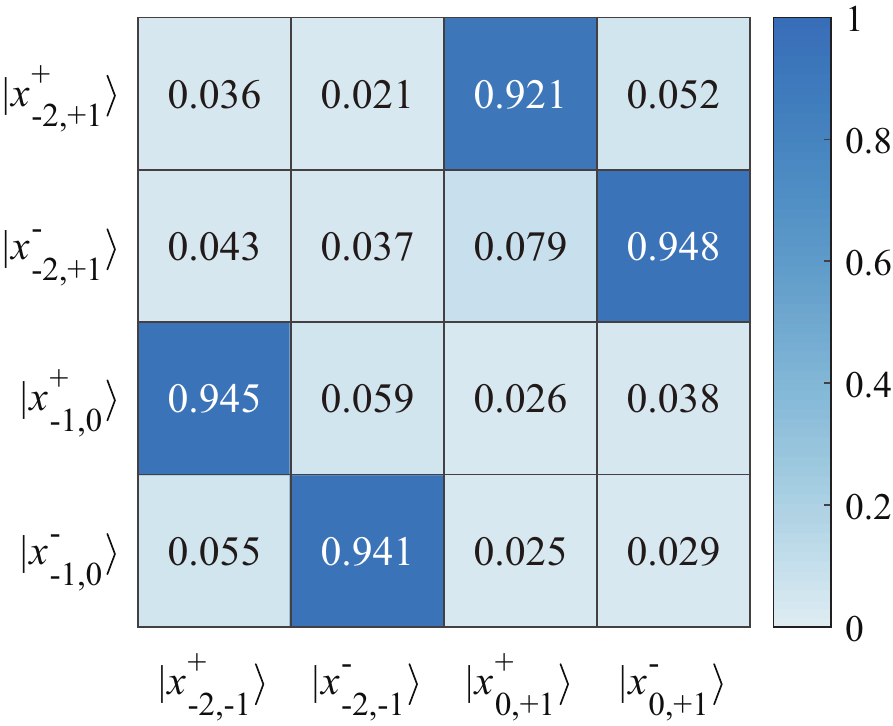}
  \caption{The measured conversion rates of the $X$ gate with different input states.  Each column gives the measured normalized coincidence rate in all possible output states for a given input state.}
  \label{CX1}
\end{figure}
\begin{figure*}[!t]
  \centering
  \includegraphics[width=0.9\linewidth]{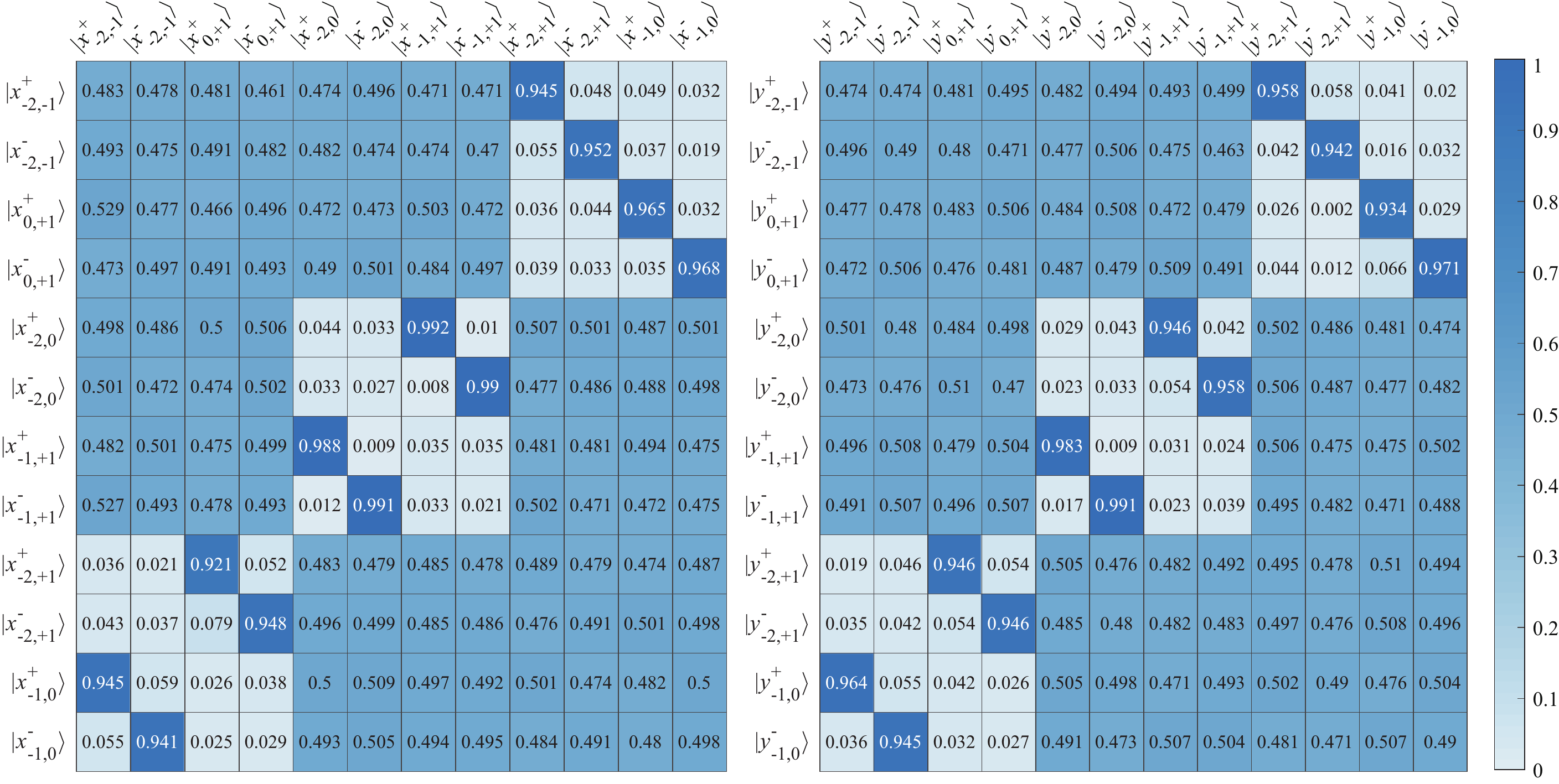}
  \caption{The outcome of the four-dimensional X gate under different seven bases. The left and right figures show the output result corresponding to the eigenstate of the operator $\sigma _x$ and  $\sigma _y$with different OAM values, respectively.  Each column gives the measured coincidence rate in all possible output states for a given input state. }
  \label{outcome2}
\end{figure*}
The $d$-dimensional Pauli $X$ gate, also namely the cyclic operation, shifts the basis state to the next following state. The integral power of the $X$ operation will transform each state to corresponding state in terms of its order. 
The generalized $Z$ gate, which applies a unique mode-dependent phase shift to each of the $ d $-dimensional base states, is given in the following matrix form
\begin{eqnarray} 
Z = \left( {\begin{array}{*{20}{c}}
1&0&0& \cdots &0\\
0&\omega &0& \cdots &0\\
0&0&{{\omega ^2}}& \cdots &0\\
 \vdots & \vdots & \vdots & \ddots &0\\
0&0&0& \cdots &{{\omega ^{d - 1}}}
\end{array}} \right),
\end{eqnarray}
where $\omega $ represents the $d^{th}$ root of unity ${e^{i2\pi /d}}$.
In this work, the OAM states are encoded in the discrete transverse LG modes of light (denoted by $\left | \ell \right\rangle, \ell\in \{-2,-1,0,1\}$), and constitute a four dimensional Hilbert space for qudit. The generalized $Z$ gate can be easily executed with a Dove prism (DP) on the OAM DoF. The Pauli matrix $Y$ for a qudit is constructed by combining $X$ and $Z$ gates. Thus the key issue for the realization of arbitrary unitary operations on a single qudit is to implement the X gate.

The experimental setup is shown in Fig.~\ref{setup}. The heralded single photon is produced via the process of type-\Rmnum{2} spontaneous parametric down-conversion process in a 5-mm-long collinear periodically poled potassium titanyl phosphate (ppKTP) crystal pumped by a 405 nm diode laser. The polarization correlated photon pairs are separated by a polarization beam splitter (PBS). The idler heralding photon with vertical polarization is coupled to a single mode fiber as trigger, when the signal photon with horizontal polarization and Gaussian transverse mode is filtered by a polarization-maintaining fiber (PMF), and then is incident on the SLM1.
The SLM1 loaded with computer-generated holograms converts the state of the signal photon with spatial Gaussian mode $\ket{ \psi_{0}}= \ket{\ell=0} $ to an arbitrary OAM quantum state $\ket{\psi_{1}}=\alpha \ket{-2}+\beta \ket{-1}+\gamma \ket{0}+\delta \ket{1} $ in which the complex amplitudes satisfy the normalization condition $\abs{\alpha}^{2} +\abs{\beta}^{2}+\abs{\gamma}^{2}+\abs{\delta}^{2}=1$.

\begin{table}[b]
\centering
\caption {The in-out conversion efficiency of the cyclic transformations with our setup.}\label{eff}
\begin{ruledtabular}
\begin{tabular}{ccccc}
Testing mode  & $ \ket{-2} $ & $ \ket{-1} $ & $ \ket{0} $ & $ \ket{1} $ \\
\hline
$ X $ gate & 91.50\% & 93.16\% & 92.19\% & 97.76\% \\
$ X^{2}  $ gate & 90.85\% & 95.56\%  & 92.68\%  &  94.80\%  \\
$  X^{\dagger} $ gate & 92.79\% & 92.07\% & 95.62\% & 91.06\% \\
\end{tabular}
\end{ruledtabular}
\end{table}

The $X$-gate operation consists of two parity splitters and a modified Mach-Zenhder interferometer in Fig.~\ref{setup}. To perform the $X$ transformation, as shown in Fig.~\ref{setup}(a), a spiral phase plate (SPP) is used to add an OAM value $+1$ and a half wave plate at 22.5$^{\circ}$ is used to change the horizontal polarization to the diagonal before the OAM parity sorter.

The OAM parity sorter is an embedded polarization-dependent Sagnac interferometer, which is designed to separate photons based on the parity of OAM values.
The DP inside is mounted at 45$^{\circ}$ so that the relative angle between the clockwise and counterclockwise paths of the interferometer effectively rotates the phase front of the incoming photon with odd or even OAM values by $\pi $ or $2 \pi $, respectively. Outside the Sagnac interferometer, as shown in Fig.~\ref{setup}(a), the photon first passes through a wave plate and then is reflected or transmitted by the PBS if its OAM value is even or odd.
The even spatial modes undergo an extra reflection which flips the sign of the OAM mode. The structure of the OAM combiner shown in Fig. \ref{setup}(a)  is opposite to that of the OAM parity sorter, which can recombine the separated odd and even OAM modes coherently.
Since the OAM combiner is used in our experiment, the cyclic operation can be accomplished without half of the photon loss caused by the stability in the previous work \cite{babazadeh2017high}.
The $X^2$ and $X^{\dagger}$ gates can be accomplished  in a similar way and are depicted in Fig.~\ref{setup}(b) and (c), respectively.

The signal photon after the $X$ gate is analyzed by the detection block consisting of the SLM2 and a single mode fiber (SMF). Here, the SLM2 flattens the phase of an incoming photon, transforming it into a Gaussian mode which can be efficiently coupled to the SMF. In this case, the OAM content of single photons can be measured for either specific modes or superposition mode. The final coincidence count for correlated photon pairs is 9.76 kHz/mW in average while the pump power is 6 mW, the accidental counts are only 5 Hz, and the coincidence efficiency is about 23.2\%. All the measurements with different OAM states are detected in one second.

In order to evaluate the quality of the cyclic transformations, we define the efficiency as
$ P\left( {i,j} \right) =  N_{ij}/\sum_{k}N_{ik},$
which denotes the probability for detecting a photon in mode $ j $ if the incoming photon in mode $ i $.  Here, $N_{ij}$ and $N_{ik}$ are the  coincidence photon counts when the input photon is
in mode $ i $ and the output one in mode $j$ and $k$, respectively, and
the summation $\sum_k$ is over all possible output modes.
The efficiency of the cyclic transformations for some typical input states is listed in Table.~\ref{eff}. We can obtain that the average conversion probabilities of the expected modes for $X$, $X^{2}$, and $X^{\dagger}$ gates are 93.66\%, 93.47\%, and 92.89\%, respectively. The main factor limiting the in-out conversion efficiency in this scheme is the phase modulation of the photon orbital angular momentum mode which is not perfect during the projection measurements. On the other hand, The reduce of the efficiency is caused by the cascade of interferometers and misalignment of optical devices. For instance, the lower detection probability for the $X^2$ gate in the input mode $ \ket{-2} $ stem mainly from spatial misalignments which led to a lower coupling efficiency for this specific mode to the single photon detector.

To further analyze the characteristic of these cyclic transformations, we consider the following eigenstates of Pauli matrices
$\sigma_x$ and $\sigma_y$, which are defined as
\begin{align}
    &\left|x_{\ell_{1},\ell_{2}}^{\pm}\right>=\frac{1}{\sqrt{2}}(\left|\ell_{1}\right>\pm\left|\ell_{2}\right>),\notag\\
    &\left|y_{\ell_{1},\ell_{2}}^{\pm}\right>=\frac{1}{\sqrt{2}}(\left|\ell_{1}\right>\pm i\left|\ell_{2}\right>).
\end{align}
In these four-dimensional OAM space, the bases are
\begin{align}
&B_2=\{\left|x_{-2,-1}^{+}\right>,\left|x_{-2,-1}^{-}\right>,\left|x_{0,+1}^{+}\right>,\left|x_{0,+1}^{-}\right>\},\notag\\
&B_3=\{\left|y_{-2,-1}^{+}\right>,\left|y_{-2,-1}^{-}\right>,\left|y_{0,+1}^{+}\right>,\left|y_{0,+1}^{-}\right>\},\notag\\
&B_4=\{\left|x_{-2,0}^{+}\right>,\left|x_{-2,0}^{-}\right>,\left|x_{-1,+1}^{+}\right>,\left|x_{-1,+1}^{-}\right>\},\notag\\
&B_5=\{\left|y_{-2,0}^{+}\right>,\left|y_{-2,0}^{-}\right>,\left|y_{-1,+1}^{+}\right>,\left|y_{-1,+1}^{-}\right>\},\notag\\
&B_6=\{\left|x_{-2,+1}^{+}\right>,\left|x_{-2,+1}^{-}\right>,\left|x_{-1,0}^{+}\right>,\left|x_{-1,0}^{-}\right>\},\notag\\
&B_7=\{\left|y_{-2,+1}^{+}\right>,\left|y_{-2,+1}^{-}\right>,\left|y_{-1,0}^{+}\right>,\left|y_{-1,0}^{-}\right>\}.
\end{align}
The three gates are examined in different OAM superposition subspaces, for instance, we select the superposition states of different OAM modes as the input target states in the basis $B_2$ for the $X$ gate, and the outcome is given in Fig.~\ref{CX1}.  In addition, all the above-mentioned superposition bases are tested in our setup, the output results are shown in Fig.~\ref{outcome2}. 
The input superposition state can maintain its coherence well after passing through the $X$ gate.

\section{High-dimensional Controlled Gates}
In addition to the unitary operations on single qudits, quantum computation based on qudits needs controlled unitary transformations as well.
Moreover, for multi-particle high-dimensional quantum computing, the interaction between particles performed through controlled gates is significantly important and can accomplish a complete high-dimensional quantum circuit.
However, it is extremely difficult to realize a double-qudit control gate since operations of $d^2\times d^2$ unitary matrices are involved.
It is a step to the realization of control qudit logic gates and to further expand the application of the high-dimensional quantum logic gates realized in our experiment.

\begin{figure*}[t]
  \centering
  \includegraphics[width=0.9\linewidth]{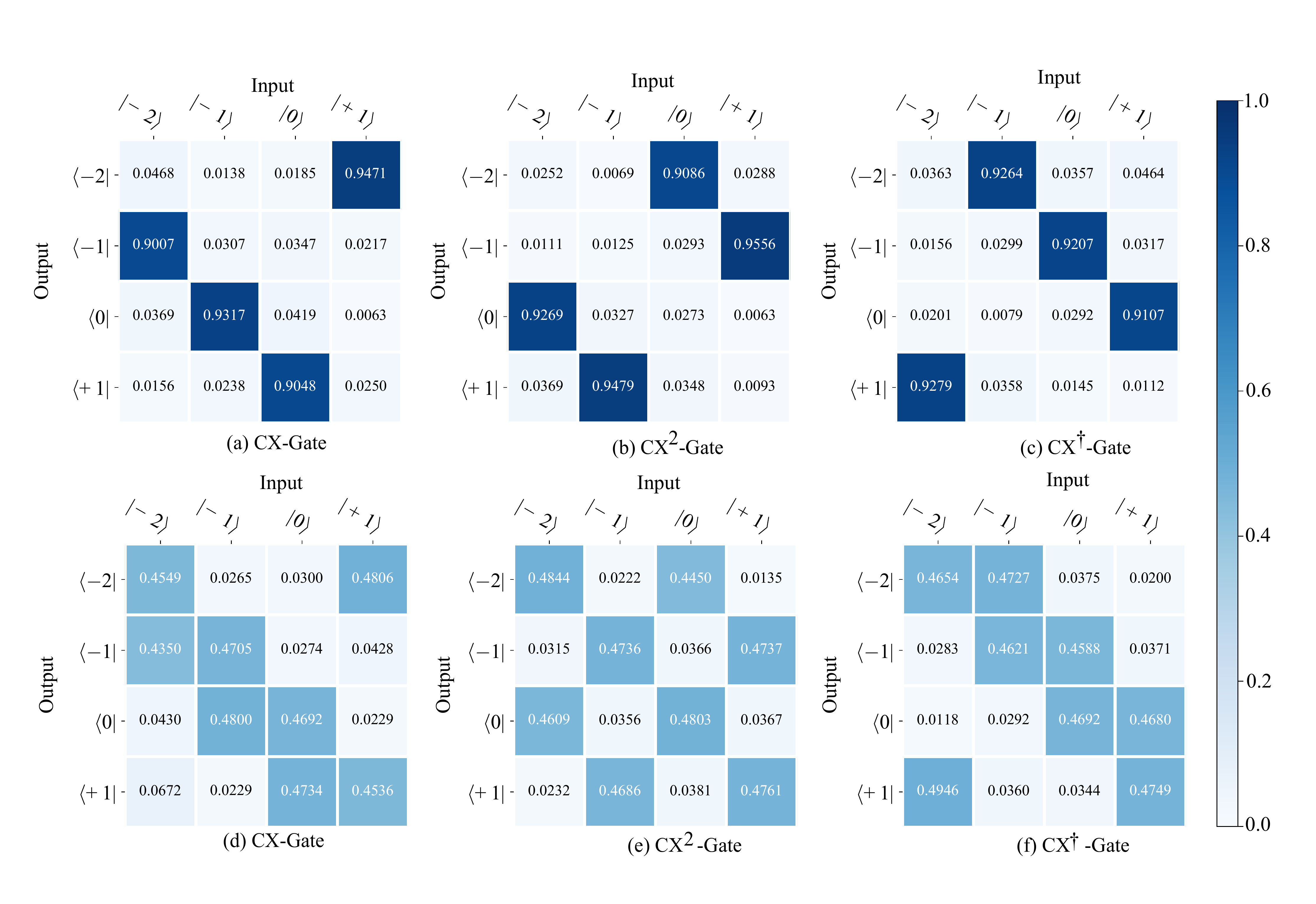}
  \caption{Outputs of the controlled gates with different input states. Each column gives the measured normalized coincidence rates in possible output modes for a given input mode. The figures (a), (b) and (c) show the performance of the three controlled gates when the control qubit is horizontally polarized. The figures (d), (e), and (f) show the  performance when the polarization of the control qubit is diagonal. }
  \label{CX}
\end{figure*}

In our experiment, we use two polarization states of a photon as control bit, which are eigenstates $\left| H \right\rangle$ and $\left|V\right\rangle$ of two-dimensional Pauli matrix $\sigma_z$. Taking the polarization DoF as controlled qubit and the OAM DoF as target states, we have realized a polarization-controlled four-dimensional $X$ gate and its integer powers at single-photon level. Together with the polarization DoF,  the control $X ^{n} $ gate in a hybrid $2d$-dimensional Hilbert space is given as
\begin{eqnarray}
C{X^n} = \left|  H  \right\rangle \left\langle  H  \right| \otimes {X^n} + \left|  V  \right\rangle \left\langle  V  \right| \otimes I_d,
\end{eqnarray}
where $I_d$ is a $ d $-dimensional identity operation. In the representation spanned by computation base, the controlled X and its integer powers can be written as $ 2d \times  2d$ matrices, i.e.
\begin{eqnarray}
CX = \left( {\begin{array}{*{20}{c}}
   {{I_d}} & 0  \\
   0 & {{X}}  \\
\end{array}} \right),C{X^n} = \left( {\begin{array}{*{20}{c}}
   {{I_d}} & 0  \\
   0 & {X^n}  \\
\end{array}} \right).
\end{eqnarray}

\begin{figure}[!t]
  \centering
  \includegraphics[width=0.9\linewidth]{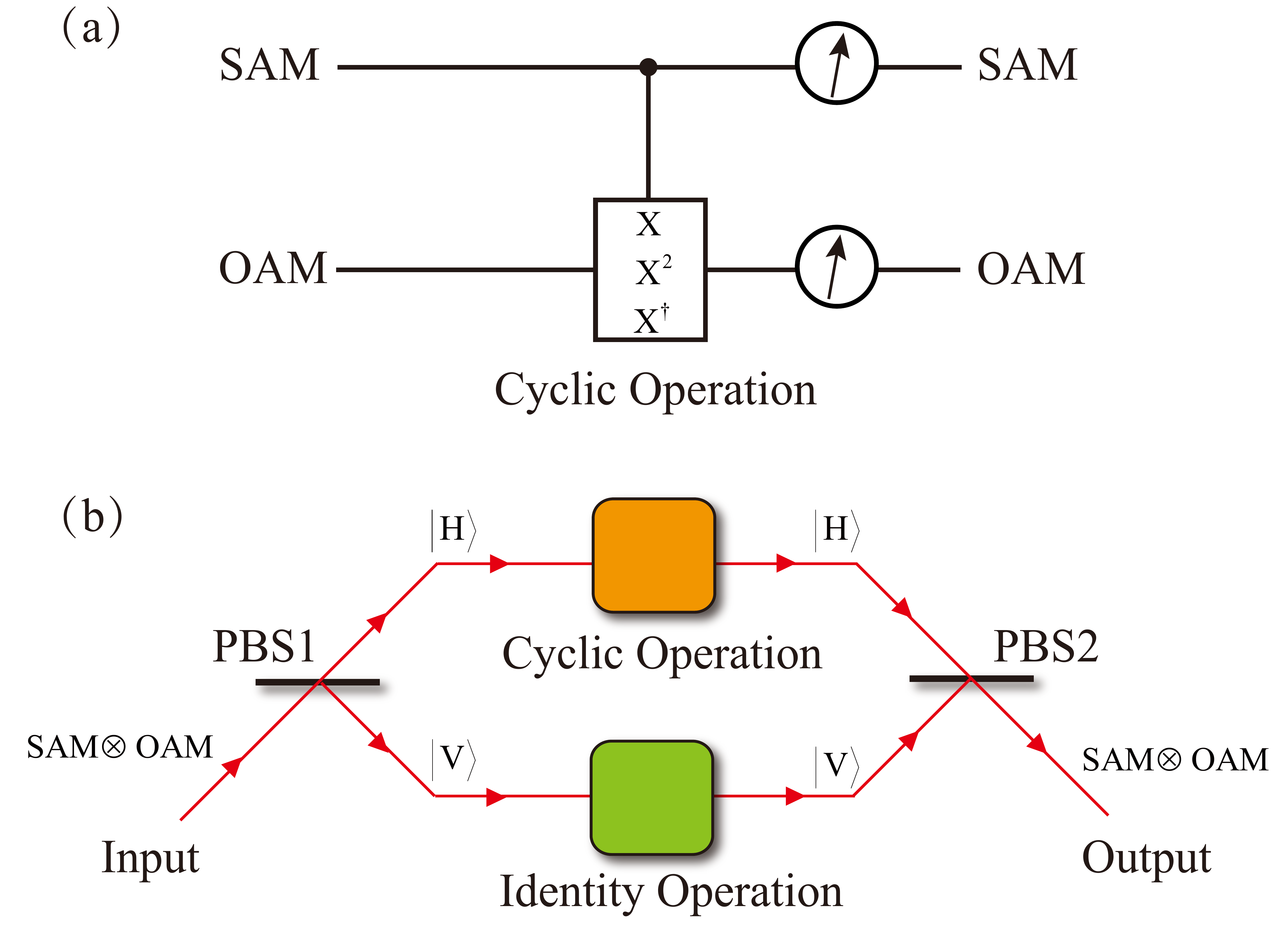}
  \caption{(a) Quantum circuit of the controlled $X$,$X^2$ and $X^\dag$. (b) Experimental scheme for the four-dimensional controlled gates. The input state of the controlled gates is the direct product state of photon's SAM and OAM.}
  \label{cgate}
\end{figure}

As shown in the quantum circuit in Fig.~\ref{cgate}(a), these gates implement the following unitary transformations on the target qubit dependent on the control qubit state. If the control qubit is in the state $\left| H\right\rangle$, a cyclic transformation operation is performed on the target state, otherwise the target state remains unchanged. The experimental scheme of the controlled gates is illustrated as in Fig.~\ref{cgate}(b).
The horizontally polarized photons pass through the PBS1 and then undergo a cyclic operation when  the OAM state of the vertically polarized photons remains the same. The photons of different polarizations are coherently recombined together at PBS2. The polarization state of the photon is detected by a quarter-wave plate, a half-wave plate and a PBS, and the projective measurements of the OAM state is achieved by loading various gratings on the SLM2.

Here we check the case where the control qubit is in either the horizontal or superposition polarization state, respectively, and the input state of the target qudit is the computational basis ($B_1$). The outcomes of the control gates are displayed in Fig.~\ref{CX}. When the polarization of the control bit is horizontal, the results in the first row of the figure show the conversion ratio of the corresponding target bit under three different cyclic operations. It can be seen that the lowest conversion efficiency can reach more than 90$\%$. In order to test the performance of the logic gates when the control bit is in the superposition state, the results in the second row of Fig.~\ref{CX} show the conversion efficiency for the corresponding target state when the control qubit is diagonally polarized.  The output results are in line with expectations, which also verifies the reliability of the logic gates in the experiment. For the four-dimensional case, the controlled $X$, $X^{2} $, and $ X^{\dagger}$ unitary transformations in combination with controlled $Z$, $Z^{2} $ and $ Z^{\dagger} $, are sufficient for the construction of arbitrary control operations on a single photon in a four-dimensional state space.


To stabilize the OAM sorter and combiner, we set the distance between the devices in the interferometer as close as possible to reduce the path length of the interferometer. In addition, the constant temperature of the laboratory reduces the influence of temperature on the stability of the interferometer as well. All of the data listed here are taken over one day. During date collection period, we only change the angles of HWPs and QWP after and before two SLMs and grating images loaded on SLMs for measurement, and its performance is repeatable.

\section{Conclusion}
In this investigation,  using the OAM beam combiner and optimizing the input mode generated after the SLM, we improve the photon conversion rate of the cyclic transformation to $ 93\% $ without loss-in-principle of photons.
In addition, we combine with the polarization state of the photons to realize a set of the high-dimensional control cyclic operations in four-dimensional OAM Hilbert space.
These high dimensional quantum logic gates have many applications for high dimensional quantum information processing such as high-dimensional quantum key distribution where transformations between mutually unbiased bases are need \cite{mafu2013higher}.
A high-dimensional generalization of the CNOT gate consisting of a controlled-cyclic transformation is necessary for a high-dimensional SWAP gate \cite{garcia2013swap}. In a polarization-path-OAM hybrid space, one can create a genuine lossless high-dimensional generalization SUM gate based on our method. Moreover, we note that a quantum state transfer scheme can also be implemented with these qubit-quart control cyclic transformations \cite{feng2020teleporting}. A next important step is the construction of high-dimensional two-particle gates, which would enable complex quantum algorithms, such as quantum error correction and other quantum algorithms\cite{bocharov2017factoring}.

\section*{Acknowledgment}

This work was supported by the National Nature Science Foundation of China (Grant Nos. 11804271, 11534008, 91736104 and 12074307) and China Postdoctoral Science Foundation via Project No. 2020M673366.

\nocite{*}
%

\end{document}